\documentclass[prl,reprint,citeautoscript,floatfix]{revtex4-2}

\usepackage{amsmath}
\usepackage{amssymb}
\usepackage{graphicx}
\usepackage{hyperref}
\begin{document}
\title{Period tripling due to parametric down-conversion in circuit QED}

\author{Lisa Arndt}
\email{lisa.arndt@rwth-aachen.de}
\affiliation{JARA Institute for Quantum Information, RWTH Aachen University, 52056 Aachen, Germany}

\author{Fabian Hassler}
\affiliation{JARA Institute for Quantum Information, RWTH Aachen University, 52056 Aachen, Germany}

\date{May 2022}
\begin{abstract}
Discrete time-translation symmetry breaking can be observed in periodically-driven systems oscillating at a fraction of the frequency of the driving force. However, with the exception of the parametric instability in period-doubling, multi-periodic driving does not lead to an instability threshold. In this paper, we point out that quantum vacuum fluctuations can be generically employed to induce period multiplication. In particular, we discuss the period-tripled states in circuit QED and propose a microwave setup. We show that for weak dissipation or strong driving, the system exhibits a non-equilibrium phase transition in the sense that the time scale over which the period-tripled state is generated can be arbitrarily separated from the time-scale of the subsequent dephasing. 
\end{abstract}
\maketitle

In a periodically-driven, nonlinear system, discrete time-symmetry breaking can be observed in the spontaneous emergence of phase-locked oscillations at an integer multiple of the driving period. This property has been attracting much attention recently in the context of discrete time crystals \cite{khemani:16,else:16,yao:17,heugel:19,kyprianidis:21}, named in analogy with charge density waves in crystalline structures. In addition, the discrete periodicity in phase space induced by multi-photon processes offers interesting possibilities to engineer tunable energy band structures \cite{guo:13,guo:16,svensson:18,guo:20,lang:21} as well as study higher-order squeezing and multipartite entanglement \cite{braunstein:87,armour:13,chang:20}.

The best-known example is the period doubling in a parametrically driven
oscillator \cite{guckenheimer,strogatz:00,wustmann:19}. There, the system
displays an instability threshold as a function of the driving strength, at
which the system undergoes a pitchfork bifurcation. The bifurcation renders
the symmetric vacuum state unstable, splitting it into two
continuously-emerging, symmetry-broken states \cite{strogatz:00,arndt:21}.
This instability threshold is an identifying feature of period doubling that
does not translate to multiple-period transitions, where instead the symmetric
vacuum state remains stable throughout. In this paper, we propose that---despite the absence of a classical instability threshold---quantum vacuum fluctuations can be used to induce period multiplication.
We analyze the time scale over
which the time-symmetry breaking occurs. We show that in the presence of weak dissipation or strong driving the timescale of the symmetry breaking is well-separated from the one of the subsequent dephasing. While our approach is valid for all
 multiple-period transitions, we discuss explicit results for the textbook example of 
period-tripling \cite{arnold:89,wustmann:19}. This serves as an generic example and 
 has been studied previously, among others, for a classical system with
thermal fluctuations \cite{tadokoro:20} as well as in the adiabatic regime
\cite{lorch:19}.

While the initial multiple-period transition is generic, the properties of the symmetry-broken states depend on the employed stabilizing potential. Transitions between the symmetry-broken states have been previously discussed \cite{zhang:17,zhang:19,gosner:20,nathan:20,lang:21} mainly using a Duffing oscillator as a stabilizing potential. Since circuit QED setups have been successfully employed to observe period multiplication \cite{denisenko:16,svensson:17,svensson:18,chang:20} in the past, we describe our procedure by means of a microwave setup in the form of a dc-biased Josephson junction coupled to a resonator that implements the studied multi-periodic transition. Here, the parametric drive is realized by the ac-Josephson effect. Furthermore, the Josephson nonlinearity leads to an interesting 6-fold symmetry in the phase of the period-tripled states---in contrast to the Duffing oscillator which leads to a 3-fold symmetry. We study the dephasing of the phase-locked, period-tripled states with  a focus on this characteristic 6-fold symmetry. This symmetry is weakly broken to a 3-fold symmetry in the presence of small dissipation and detuning.

The article is organized as follows: after introducing the proposed microwave setup, we describe the system with a time-dependent Hamiltonian. We perform a rotating-wave approximation and discuss how the symmetric vacuum state can be turned unstable. Employing a quasi-classical Martin-Siggia-Rose action valid for small impedance, we investigate how dissipation influences the proposed phase-transition. Finally, we analyze the 6-fold symmetry breaking in the limit of small detuning and dissipation.

%
\begin{figure}[tb]
	\centering
	\includegraphics{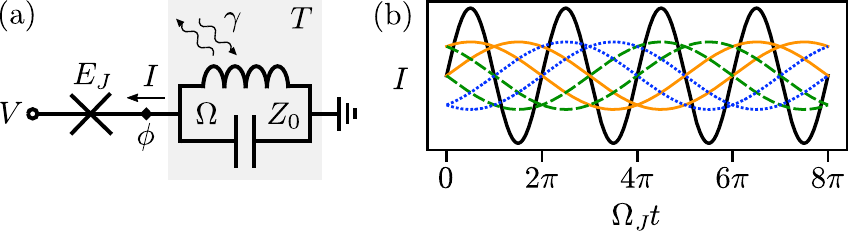}
	\caption{%
		(color online) (a) Setup composed of a Josephson junction with Josephson energy $E_J$ biased by a dc voltage $V$ and coupled to a microwave resonator characterized by its resonance frequency $\Omega$ and an impedance $Z_0$. The resonator is coupled to an environment at temperature $T$ with a coupling rate $\gamma$. The superconducting phase difference across the junction depends on the superconducting phase $\phi$ across the resonator as well as the applied dc-voltage $V$. (b) Oscillating current $I$ as indicated in (a) in arbitrary units for the system in the vacuum state (black, solid) as well as for the 6 phase-locked period-tripled states (colored, thin) in the limit of small dissipation and zero detuning. At larger dissipation, first the solutions indicated by the same style mix before the system dephases completely.  
	}\label{fig:setup}
\end{figure}
The proposed setup for the observation of period multiplication is illustrated in Fig.~\ref{fig:setup}(a). It is composed of a Josephson junction with Josephson energy $E_J$ that is biased by a dc-voltage source and in series with a microwave resonator. The resonator is characterized by its resonance frequency $\Omega$ and an impedance $Z_0$ at low frequency. The resonator is coupled to an environment at temperature $T$ with a small coupling rate $\gamma\ll\Omega$. Close to resonance, the impedance $Z(\omega)$ of the resonator is given by
\begin{equation}
Z(\omega)=\frac{Z_0\Omega}{\gamma-i(\omega-\Omega)},
\end{equation}
valid for $|\omega-\Omega|\ll\Omega$.  The superconducting phase difference across the junction depends on the superconducting phase $\phi$ across the resonator as well as the applied dc-voltage $V$. To observe $m$-fold period multiplication, we tune the Josephson frequency $\Omega_J$ close to $m$ times the resonance frequency of the microwave resonator by setting the dc-bias voltage to $V=\hbar\Omega_J/2e$, with $\Omega_J=m(\Omega-\Delta)$. Here, the detuning $\Delta$ is assumed to be small with $|\Delta|\ll\Omega$. 


In the absence of dissipation (with $\gamma=0$), the quantum system can be described by the time-dependent Hamiltonian
\begin{align}\label{eq:ham}
H=\hbar\Omega a^\dag a-E_J\cos\left[\Omega_J t-\tfrac12\sqrt{\kappa}\left(a^\dag+a\right)\right].
\end{align}
Here, $a$ and $a^\dag$ correspond to the conventional ladder operators with the commutation relation $[a,a^\dag]=1$. The ladder operators are related to the superconducting phase via $\phi=\frac12\sqrt{\kappa}\left(a^\dag+a\right)$ with the vacuum fluctuation strength $\kappa=16 \pi Z_0G_Q$ and the conductance quantum $G_Q=e^2/\pi\hbar$. In the following, we are interested in the quasi-classical limit of small vacuum fluctuation strength $\kappa\ll1$. In this regime, the impedance $Z_0$ far from resonance is small on a quantum scale such that contributions to the phase fluctuation away from the resonance frequency can be neglected. 

To study the system further, we move to a rotating frame via the transformation $U(t)=\exp[i\Omega_J a^\dag a t/m]$ and perform a rotating-wave approximation to obtain a time-independent Hamiltonian. Close to resonance with $|\Delta|\ll\Omega$, this Hamiltonian describes the system well, provided the vacuum fluctuation strength $\kappa$ and the coupling strength $E_J$ are not too strong \cite{armour:13}. We obtain \cite{lang:21}
\begin{align}\label{eq:hamrwa}
\tilde{H}=\hbar\Delta a^\dag a-\frac{\tilde{E}_J}{2}
\colon\!\!\frac{J_m\Bigl(\!\sqrt{\kappa a^\dag a}\Bigr)}{(a^\dag a)^{m/2}}
\Bigl[(-ia)^m\!+\!(ia^\dag)^m      \Bigr]\colon
\end{align}
with $J_{m}$ the Bessel function of the first kind. The colons signify normal-ordering of the ladder operators. The Josephson energy is renormalized by the vacuum fluctuation strength with $\tilde{E}_J=E_J e^{-\kappa/8}\approx E_J$. To increase readability, we focus on the case $\Delta\geq 0$ in the following. 

For the symmetry-breaking transition from the vacuum state at the origin, only the leading-order contribution of the Bessel function $J_m(x)\approx x^m/ m!\,2^m$ is relevant, leading to the generic Hamiltonian $\tilde{H}\approx\hbar\Delta a^\dag a-\hbar\epsilon\kappa^{m/2-1}[(-ia)^m\!+\!(ia^\dag)^m]/m$ with $\epsilon=\kappa E_J/(m-1)!\,2^{m+1}\hbar$ that is independent of the stabilizing potential \cite{arnold:89,wustmann:19}. From this Hamiltonian, it is immediately evident why period doubling stands out from all higher-order multi-periodic transitions: For $m=2$, the parametric drive acts with the same power of $a$ as the detuning (and as discussed later the dissipation). As a result, the parametric drive influences the linear dynamics around $a=0$ and can, thus, change the stability of the vacuum state. This is no longer the case for $m\geq 3$, where instead the parametric drive has no impact on the stability matrix at the origin. It can, however, influence the properties of the basin of attraction around the vacuum state. This property is employed below to induce multi-periodic instabilities.

\begin{figure}[tb]
	\centering
	\includegraphics{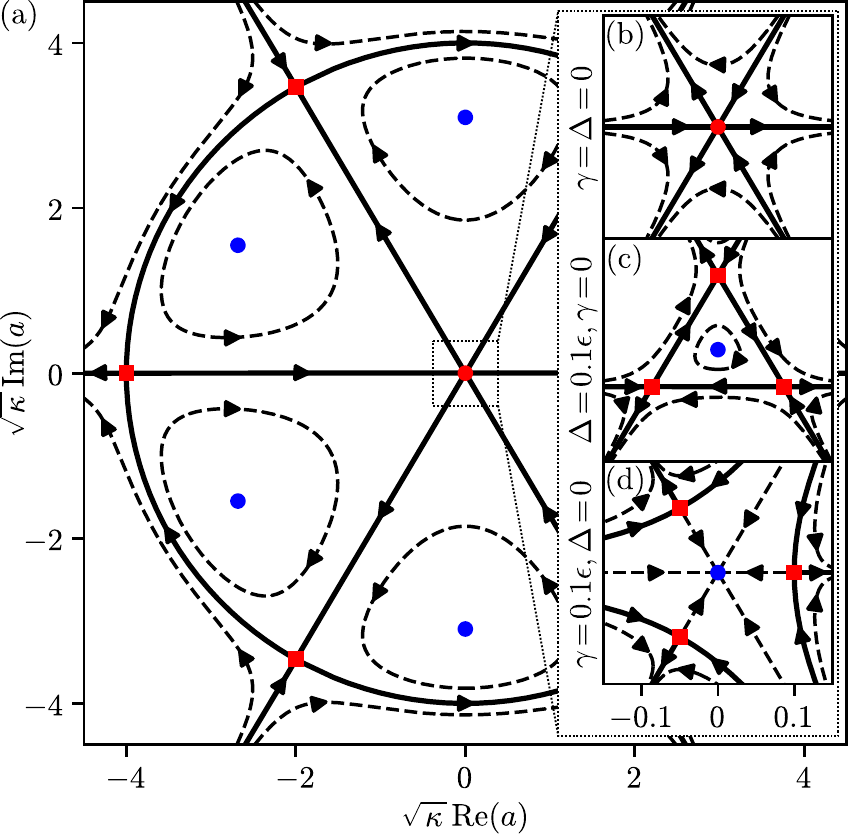}
	\caption{%
		(color online) (a) Phase portrait of $\tilde{H}$ in Eq.~\eqref{eq:hamrwa} for $m=3$ and $\Delta=0$. The system features $6$ stable fixed-points (circle, blue), $6$ unstable fixed-points (square, red), as well as one effectively unstable fixed-point at $\alpha=0$ which exhibits a vanishing stability matrix (magnified in (b)). The separatrices (black, solid) are the boundaries of the basin of attraction for each of the stable fixed-points. (c) Zoomed-in phase portrait of $\tilde{H}$ for $\Delta=0.1\epsilon$. The effectively unstable fixed-point at zero detuning is split up into a stable vacuum state as well as three equidistant unstable fixed-points forming a small, triangular basin of attraction. (d) Zoomed-in phase portrait of Eq.~\eqref{eq:MSR} in the absence of fluctuations for $\Delta=0$ and $\gamma=0.1\epsilon$. Similar to the effect of the detuning, a finite dissipation causes a split up of the effectively unstable fixed-point at the origin.
	}\label{fig:contour}
\end{figure}

The phase portrait of $\tilde{H}$ for the complex amplitude $\alpha=\langle a\rangle$ in the leading-order approximation is illustrated in Fig.~\ref{fig:contour}(b) and (c) for the period-tripling case ($m=3$). An important feature is the fixed-point at the origin in the absence of detuning, illustrated in Fig.~\ref{fig:contour}(b), which exhibits a vanishing stability matrix. In the presence of a small detuning $\Delta\ll\epsilon$, this fixed point is split up into a stable vacuum state at the origin as well as $m$ equidistant unstable fixed-points at $\alpha=i(\Delta/\epsilon)^{1/(m-2)}e^{2\pi i n/m}/\sqrt{\kappa}$ with $n=0,1,\dots,m-1$ as demonstrated in Fig.~\ref{fig:contour}(c). The $m$ unstable fixed-points form a small, polygonal basin of attraction around the vacuum state. As a result, a system initially prepared in the vacuum state for $\Delta\neq 0$ can be forced to undergo a spontaneous symmetry breaking to a multi-periodic state if the detuning is tuned to zero. Indeed, we show below that it suffices to reach the regime $(\Delta/\epsilon)^{1/(m-2)}\ll\sqrt{\kappa}$ where quantum fluctuations are sufficiently strong to cause an escape from the vacuum state. In the next section, we discuss how dissipation to a thermal reservoir influences this transition.

%


The coupling of the microwave resonator to a thermal reservoir leads to a decay of the excitations and to noise. In the presence of small dissipation $\gamma\ll\Omega$ as well as small vacuum fluctuations $\kappa\ll1$, the system can be described by the quasi-classical Martin-Siggia-Rose (MSR) action $S_\mathrm{MSR}=i\int dt[ \overline{\alpha_q}\dot{\alpha}-\alpha_q \dot{\overline{\alpha}}+\Lambda(\alpha,\alpha_q)]$ with \cite{schmid:82,kamenev_2011} 
\begin{align}\label{eq:MSR}
\!\!\!\Lambda\!=\,& \overline{\alpha_q}\!\left(\!\!\frac{i}{\hbar}\!\frac{\partial \tilde{H}}{\partial \overline{\alpha}}\!+\!\gamma \alpha\!\right)\!\!+\!\alpha_q\!\!\left(\!\!\frac{i}{\hbar}\!\frac{\partial \tilde{H}}{\partial \alpha}\!-\!\gamma \overline{\alpha}\!\right)\!\!+\!\gamma(2n_\Omega\!+\!1)\overline{\alpha_q}\alpha_q.\!\!\!
\end{align}
Here, we introduced both the response field $\alpha_q$ that encodes the fluctuations of the complex amplitude $\alpha$ as well as the Bose-Einstein occupation $n_\Omega=[\exp(\hbar\Omega/k_B T)-1]^{-1}$ of the resonator. In the quantum limit $k_BT\ll\hbar \Omega$, we obtain $n_\Omega\to 0$ and $\alpha_q$ encodes the pure quantum fluctuations of $\alpha$. In the classical limit $k_BT\gg\hbar \Omega$, the Bose-Einstein occupation increases with $2n_\Omega+1\approx 2k_BT/\hbar \Omega$ and the dynamics is governed by thermal fluctuations. By employing a Hubbard-Stratonovich transformation, it can be shown that this description is equivalent to the quasi-classical Langevin equation \cite{schmid:82} $\dot{\alpha}(t)=-(i/\hbar) \partial \tilde{H}/\partial \overline{\alpha}(t)-\gamma \alpha(t)+\xi(t)$. Here, the fluctuations are no longer encoded by $\alpha_q$, but instead via the complex white noise source $\xi(t)$ with the correlator $\langle\!\langle\xi(t)\overline{\xi(t')}\rangle\!\rangle=\gamma(2n_\Omega+1)\delta(t-t')$. 

The phase portrait of the period-tripling case in the presence of dissipation is illustrated in Fig.~\ref{fig:contour}(d). Similarly to the detuning, a small dissipation $\gamma\ll\epsilon$ causes the split up of the effectively unstable fixed-point into a stable vacuum state as well as $m$ equidistant unstable fixed-points. In the presence of both detuning and dissipation, these points are positioned at $\sqrt{\kappa}|\alpha|=[(\Delta^2+\gamma^2)/\epsilon^2]^{1/(2m-4)}$. It follows that in the limit $\Delta\to 0$ these points no longer converge to the origin. Instead, we require that $(\gamma/\epsilon)^{1/(m-2)}\ll\sqrt{\kappa}$ for quantum fluctuations to induce the desired symmetry-breaking transition. In the next step, we employ the MSR action to further investigate this claim and obtain explicit results of the tunneling rate \cite{Note1} in the case of period tripling.

The construction of the MSR action guarantees that $\Lambda$ is a conserved quantity. It takes the place of a Hamiltonian for the dissipative motion. A special role is played by the trajectories for $\Lambda=0$ as the fixed points as well as all other trajectories with $\alpha_q=0$ lie on this value of $\Lambda$. The saddle-point equations $\partial S_\mathrm{MSR}/\partial \alpha_q$ and $\partial S_\mathrm{MSR}/\partial \alpha$ for $\alpha_q=0$ correspond to the noiseless Langevin equation describing the classical relaxation process. Note that the action along the corresponding trajectories is always zero \cite{kamenev_2011}.

In contrast, trajectories at a finite value of $\alpha_q$ can only take place in the presence of fluctuations. It follows that in order to escape from the basin of attraction of the vacuum state, the system needs to follow a trajectory at finite $\alpha_q$, accumulating a finite action. In order to approximate the escape rate from the vacuum state to exponential accuracy, we need to find the most probable escape path. Specifically, we need a trajectory starting at the vacuum state that extends to or beyond one of the separatrices and minimizes the absolute value of the action. 

For weak noise, it is most probable to escape the basin of attraction near a fixed point \cite{kamenev_2011}. In particular, the least action is acquired along a trajectory with $\Lambda=0$ that connects the vacuum state with one of the unstable fixed-points. Since the escape problem is three-fold symmetric, the probability to escape via any of the three fixed points is the same. While the optimal trajectory takes an infinitely long time to traverse, the corresponding action $S_\mathrm{MSR}=i\int dt( \overline{\alpha_q}\dot{\alpha}-\alpha_q \dot{\overline{\alpha}})$ remains finite and the escape rate is given by $\Gamma\propto e^{iS_\mathrm{MSR}}$ to exponential accuracy, valid for $(2n_\Omega+1)\kappa\ll1$.

Here, we mainly focus on the limit $\Delta\to 0$, for which the system has the highest probability to switch to a symmetry-broken state. For $\Delta=0$, the trajectory is given by a straight radial line between the fixed points, see Fig.~\ref{fig:contour}(d). This line becomes curved for a finite value of $\Delta$ as the system in the rotating frame begins to oscillate around the origin. By solving the saddle-point equations perturbatively for $m=3$ and $\Delta\ll \gamma,\epsilon$, we obtain \cite{Note2}
\begin{align}\label{eq:rate}
\Gamma\propto\exp\left[-\frac{2(\gamma^2+\Delta^2)}{3(2n_\Omega+1)\kappa\epsilon^2}\right].
\end{align}
For completeness, we note that it is also possible to employ an approach based on the Poincare cross-section method \cite{chinarov:93} to obtain the escape time in the limit $\gamma\to0$. Using this method, it is straigthforward to rederive the result for $\gamma =0$ of  the perturbative approach above \cite{Note3}. 
We conclude that, for a system with weak dissipation $(\gamma/\epsilon)^{1/(m-2)}\ll\sqrt{\kappa}$, the vacuum state can be made unstable by decreasing the detuning $\Delta$ to zero. 

In the following, as an example, we investigate the specific properties of the multi-periodic state and its dephasing time in the introduced microwave setup.
For this purpose, we expand the Bessel function up to $5$th order with $J_m(x)\approx x^m/ m!\,2^m-x^{m+2}/(m+1)!\,2^{m+2}$.
The resulting phase portrait of $\tilde{H}$ for the period-tripling case is illustrated in Fig.~\ref{fig:contour}(a). After the escape from the basin of attraction of the vacuum state, the system is nearly guaranteed to end up close to one of the $2m$ outer stable fixed-points. For $\Delta=\gamma=0$, all fixed points have the same amplitude and differ in phase by $\pi/m$. For even $n$, the transformation $a\mapsto a\, e^{\pi i n/m}$ leaves the system unchanged, while for odd $n$ the transformation is equivalent to a time-reversal operation. As such the dynamics in the vicinity of the $2m$ stable fixed-points alternates between clockwise and counter-clockwise rotation. For $m=3$, the corresponding frequency is given by $24 \epsilon/5$. Furthermore, the stable fixed-points are positioned at $\alpha=i\sqrt{48/5 \kappa}\, e^{\pi i n/3}$ where $n=0,1,\dots,5$. The separatrices in Fig.~\ref{fig:contour}(a) are the boundaries of the basin of attraction to these $6$ stable states. They form a circular sector with the radius $|\alpha|=4/\sqrt{\kappa}$ and the dividing radii formed by the lines connecting the origin with the $6$ unstable fixed-points at $\alpha=4 e^{\pi i n/3}/\sqrt\kappa$.

The $2m$-fold symmetry of the fixed-points is weakly broken in the presence of small detuning and small dissipation. To first order in $\Delta/\epsilon$ and $\gamma/\epsilon$, the fixed-points of the period-tripling example are positioned at $\sqrt{\kappa}|\alpha|=\sqrt{48/5}\pm \Delta/2\epsilon$ with the corresponding complex phases $\varphi=\mp\pi/2\mp \sqrt{125/1728}\,\gamma/\epsilon+2\pi n/3$ with $n=0,1,2$. In the desired limit $\Delta\to 0$, the system is equally likely to end up close to any of the $2m$ stable fixed-points as each of the $m$ unstable fixed-points relevant for the multi-periodic transition is equally connected to two outer stable fixed-points. For $m=3$, the current in the laboratory frame oscillates approximately with $I(t)=d\langle 2e  q(t)\rangle/dt=-(4e\Omega_J/3\sqrt{\kappa})|\alpha|\cos(\Omega_J t/3-\varphi)$ with the dimensionless charge $q=i(a^\dag-a)/\sqrt{\kappa}$ and $[\phi,q]=i$. This phase-locking to the original drive $I(t)=(2eE_J/\hbar)\sin (\Omega_J t)$ in the vacuum state is illustrated in Fig.~\ref{fig:setup}(b). To properly observe the phase-locking, we require the timescale over which the period-tripling transition occurs to be much shorter than the timescale over which the phase-locked state is washed out. 

\begin{figure}[tb]
	\centering
	\includegraphics{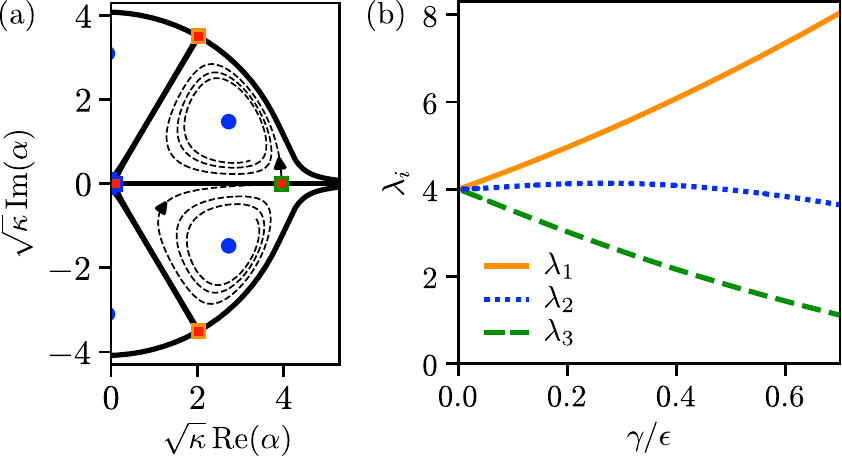}
	\caption{%
		(color online) (a) Basins of attraction of Eq.~\eqref{eq:MSR} for $m=3$, $\Delta=0$, and $\gamma=0.1\epsilon$ (black, solid). A tunneling trajectory as well as a possible relaxation trajectory are indicated by dashed lines. There are three fixed-points over which escape is possible; framed in orange, blue, and green. (b) Tunneling amplitudes (for $\Delta=0$) via the three fixed points as identified in (a). At elevated $\gamma/\epsilon$, tunneling via the green fixed point $\lambda_3$ is much more likely. As a result, the system first dephases to a three-fold symmetric state before dephasing completely.
	}\label{fig:dephasing}
\end{figure}

In order to calculate the timescale over which the broken-symmetry state dephases, we employ the MSR formalism introduced above for the period-tripling example. For each of the six outer stable fixed-points, there are three likely fixed points of escape that are shown in Fig.~\ref{fig:dephasing}(a): the unstable fixed-point close to the origin that enabled the initial symmetry breaking and the two adjacent outer unstable fixed-points. In general, each unstable fixed-point is connected to a different escape rate and we obtain $\Gamma\propto\exp[-\lambda_i(\gamma/\epsilon,\Delta/\epsilon)/(2n_\Omega+1)\kappa]$ with $i=1,2,3$ and $\lambda_i=\mathcal{O}(1)$ for $\Delta,\gamma\ll\epsilon$. These dephasing rates are much smaller than the initial period-tripling rate in Eq.~\eqref{eq:rate} which scales exponentially with $\gamma^2/\epsilon^2$. Therefore, the timescale over which the period-tripled state is generated can be arbitrary separated from the timescale of the subsequent dephasing. Thus, observation of the six possible phase-locked oscillations is feasible.
For $\Delta=0$, numerical results of $\lambda_i$ are shown in Fig.~\ref{fig:dephasing}(b). They are obtained by solving the saddle-point equations of $\Lambda$ in Eq.~\eqref{eq:MSR} backwards in time, starting from one of the three unstable fixed-points respectively and optimizing the initial parameters until the solution connects to the stable fixed-point. In the limit $\gamma/\epsilon\ll1$, the local dynamics around the stable fixed-points is strongly underdamped and the system typically performs many rotations along triangle-shaped trajectories around the stable fixed-point without a loss of local amplitude. This is reflected in the optimal escape trajectory. Since the three unstable fixed-points define the outer border of these almost closed loops, the three tunneling rates coincide for $\gamma\to0$. Employing the analytic approach based on the Poincare cross-section method \cite{chinarov:93}, we obtain $\lambda_1=\lambda_2=\lambda_3\approx 3.997$. 

At an elevated but small value of $\gamma/\epsilon$, the tunneling $\lambda_3$ via one of the two outer unstable fixed-points becomes the most likely as the distance between the stable fixed-point and this unstable fixed-point decreases linearly with $\gamma/\epsilon$. This particular fixed point only leads to the mixing of the two adjacent stable fixed-points as shown in Fig.~\ref{fig:dephasing}(a). In contrast, the distance to the other outer unstable fixed-point increases linearly with $\gamma/\epsilon$, decreasing the corresponding tunneling rate $\lambda_1$ such that tunneling via this fixed point becomes even less likely than tunneling via the inner unstable fixed-point denoted by $\lambda_2$. In general, tunneling via the inner point leads to a complete dephasing of the initial symmetry-broken state. As such, the system has the tendency to first reach a 3-fold symmetric state before dephasing completely. As apparent by the tunneling ratio $\Gamma_{\mathbb{Z}3}/\Gamma_{\mathbb{Z}6}\propto\exp[(\lambda_2-\lambda_3)/(2n_\Omega+1)\kappa]$, this effect is particularly strong at very small values of $\kappa$. 

In conclusion, we have outlined a generic method to induce multi-periodic instabilities that lead to symmetry-breaking transitions. While our explicit results focused on the period-tripling transition as a textbook example, our approach to induce a symmetry-breaking transition by decreasing the detuning between the parametric drive and the resonance frequency is valid for all higher-order transitions.  We have shown that the timescale over which the symmetry broken state is generated can be arbitrary separated from the time-scale of the subsequent dephasing. Furthermore, we have analyzed the properties of the symmetry-broken states in the context of a microwave setup based on a Josephson stabilizing potential where a $6$-fold symmetry emerges in the phase of the period-tripled states.  In the presence of weak dissipation, the system first reaches a 3-fold symmetric state before the information about the phase is lost completely. 

\begin{acknowledgments}
This work was supported by the Deutsche Forschungsgemeinschaft (DFG) under Grant No.~HA 7084/6-1.
\end{acknowledgments}

\onecolumngrid\clearpage\setcounter{equation}{0}
\renewcommand\theequation{S\arabic{equation}}
\setcounter{figure}{0}
\renewcommand\thefigure{S\arabic{figure}}

\section{Supplement}\setcounter{page}{1}
\subsection{Detailed derivation of the escape rate for the period-tripling transition}
For the following calculation, we focus on the period-tripling Hamiltonian $\tilde H\approx \hbar\Delta a^\dag a-i\hbar\epsilon\sqrt{\kappa}[a^3-(a^\dag)^3]/3 $. To simplify the corresponding MSR action described by Eq.~\eqref{eq:MSR}, we substitute $\alpha=(q_1+iq_2)/\sqrt{\kappa}$ and $\alpha_q=(p_2+ip_1)/\sqrt{\kappa}$. Furthermore, we make the simple linear change of variables $p_1\mapsto ip_1/(2n_{\Omega}+1)$ and $p_2\mapsto -i p_2/(2n_{\Omega}+1)$ such that the dissipative action acquires the form \cite{kamenev_2011}
\begin{align}
&iS_\mathrm{MSR}=-\frac{2}{\kappa (2n_\Omega+1)}\int dt \left[p_1\dot{q_1}+p_2\dot{q_2}-H_S(q_1,q_2,p_1,p_2)\right],\,\text{with the fictitious Hamiltonian}\nonumber\\ &H_S=\tfrac12\gamma (p_1^2+p_2^2)-\gamma(p_1 q_1+p_2 q_2)+\Delta (p_1 q_2-p_2 q_1)+\epsilon(p_1 q_1^2-2 p_2 q_1 q_2-p_1 q_2^2).\nonumber
\end{align}
The MSR action leads to the corresponding saddle-point equations $\partial H_S/\partial p_i=\dot q_i$ and $-\partial H_S/\partial q_i=\dot p_i$. As explained in the main text, we need to find a trajectory for $H_S=0$ with a finite value of at least one $p_i$ that connects the vacuum state at $p_i=q_i=0$ with one of the unstable fixed-points, \emph{e.g.} at $p_i=0$, $q_1=\gamma/\epsilon+4\Delta^2/9\gamma\epsilon$, and $q_2=-\Delta/3\epsilon$ to second order for $\Delta\ll\gamma,\epsilon$. Note that it is useful to rewrite the saddle-point equations as a function of $q_1$ instead of $t$. 

For $\Delta=0$, Fig.~\ref{fig:contour}(d) already implies that the trajectory is given by a straight radial line. Indeed, the remaining $3$ saddle-point equations for $q_2(q_1)$, $p_1(q_1)$, and $p_2(q_1)$ together with the restraint $H_S=0$ have the simple solution $p_1^{(0)}=2q_1(\gamma-\epsilon q_1)/\gamma$ and $q_2^{(0)}=p_2^{(0)}=0$. Integrating the action results in the escape rate $\Gamma\propto\exp(iS_\mathrm{MSR}^{(0)})= \exp[-2\gamma^2/3\epsilon^2\kappa (2n_\Omega+1)]$.

For $\Delta\neq 0$, we make the perturbative ansatz 
\begin{align}\nonumber
p_1(q_1)&=p_1^{(0)}(q_1)+(\Delta/\epsilon) \,p_1^{(1)}(q_1)+(\Delta/\epsilon)^2 p_1^{(2)}(q_1),\\\nonumber
q_2(q_1)&=(\Delta/\epsilon)\, q_2^{(1)}(q_1),\\\nonumber
p_2(q_1)&=(\Delta/\epsilon)\, p_2^{(1)}(q_1).\end{align}
Inserting this ansatz into the saddle-point equations allows for a successive solution for the first and second order in $\Delta/\epsilon$ by employing the restraint $H_S=0$ as well as the initial conditions at the fixed points listed above. This immediately yields $p_1^{(1)}=0$. For the MSR action, we obtain
\begin{align}\nonumber
&iS_\mathrm{MSR}=iS_\mathrm{MSR}^{(0)}-\frac{2\Delta^2}{\epsilon^2\kappa (2n_\Omega+1)}\int_{0}^{\gamma/\epsilon} d q_1 \Big[\underbrace{p_2^{(1)}(q_1)q_2'^{(1)}(q_1)}_{\to \frac{253}{108}-\frac{11\pi^2}{45}}\,\,+\!\!\underbrace{p_1^{(2)}(q_1)}_{\to -\frac{217}{108}+\frac{11\pi^2}{45}}\!\!\!\!\!\!\Big]=-\frac{2(\gamma^2+\Delta^2)}{3\epsilon^2\kappa (2n_\Omega+1)},
\end{align}
with the corresponding escape rate $\Gamma\propto\exp(iS_\mathrm{MSR})$ of Eq.~\eqref{eq:rate}. 

\subsection{Numerical simulation of the quasi-classical Langevin equation}

To validate the analytical result for the escape rate $\Gamma$ in Eq.~\eqref{eq:rate}, we simulate the tunneling dynamics of the corresponding period-tripling Hamiltonian $\tilde H\approx \hbar\Delta a^\dag a-i\hbar\epsilon\sqrt{\kappa}[a^3-(a^\dag)^3]/3 $ with the quasi-classical Langevin equation \cite{schmid:82} $\dot{\alpha}(t)=-(i/\hbar) \partial \tilde{H}/\partial \overline{\alpha}(t)-\gamma \alpha(t)+\xi(t)$ for the complex amplitude $\alpha$. Here, $\xi(t)$ is a complex white noise source with the correlator $\langle\!\langle\xi(t)\overline{\xi(t')}\rangle\!\rangle=\gamma(2n_\Omega+1)\delta(t-t')$. To obtain the escape rate $\Gamma\propto\exp(-R)$, we sample 3000 trajectories for each set of parameters. Every trajectory begins from a point $\alpha_0$ close to the origin that is Gaussian distributed with $\langle\alpha_0\rangle=0$ and $\langle|\alpha_0|^2\rangle=\frac12 +n_\Omega$. The trajectory is cut-off well behind the basin of attraction of the vacuum state such that the result is independent of the chosen boundary. 

\begin{figure}[b]
	\centering
	\includegraphics{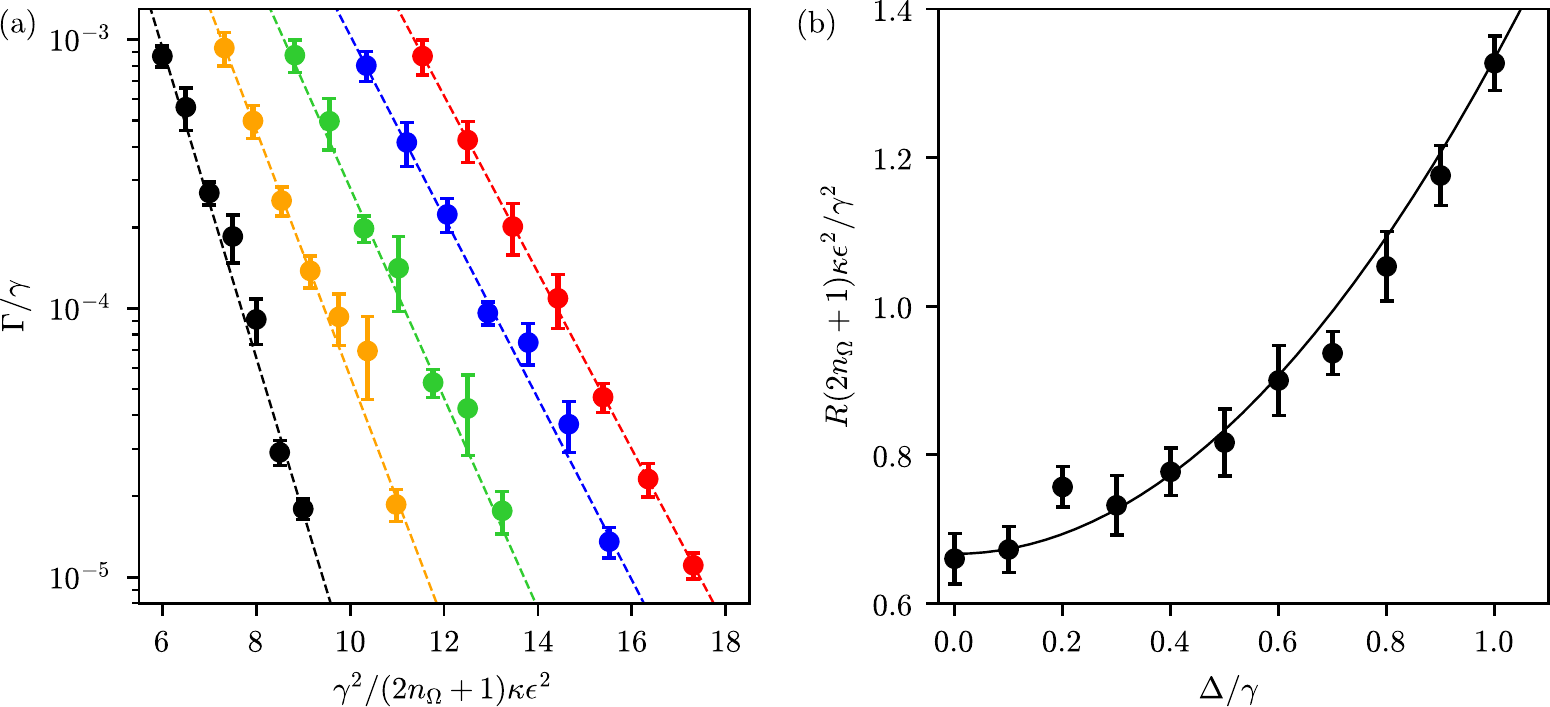}
	\caption{%
		(color online) (a) Logarithmic plot of the tunneling rate $\Gamma$ from the vacuum state as a function of the inverse fluctuation strength $1/\kappa(2n_\Omega+1)$ as obtained by numerically simulating the Langevin dynamics for exemplary values of $\Delta$. The full circles correspond to the numerical tunneling rate with the corresponding statistical errors obtained by the simulation. The dashed lines are linear fits of the numerical data that were performed to obtain the exponential factor $R$ with $\Gamma\propto\exp(-R)$. From right to left, the colors correspond to $\Delta/\gamma=0.2,0.4,0.6,0.8,1$. (b) Dependence of the exponential factor $R$ on the detuning $\Delta$. The full circles represent the values obtained by the linear fit of the values in (a) with the propagated errors. The solid line corresponds to the analytical prediction in Eq.~\eqref{eq:rate}. Both results are in good agreement.}\label{fig:sup}
\end{figure}

The results of the numerical simulation are shown in Fig.~\ref{fig:sup}. Figure~\ref{fig:sup}(a) clearly demonstrates the linear dependence of the logarithmic tunneling rate $\log (\Gamma)$ on the inverse fluctuation strength $1/\kappa(1+2n_\Omega)$ for fixed values of $\Delta/\gamma$. The fitted slopes are compared to the analytical prediction $R=2(\gamma^2+\Delta^2)/3(2n_\Omega+1)\kappa\epsilon^2$ in Fig.~\ref{fig:sup}(b). The values of the logarithmic tunneling rate obtained by simulations and from the analytical theory are in good agreement.


\begin{thebibliography}{10}
\makeatletter

\bibitem{khemani:16}
V. Khemani, A. Lazarides, R. Moessner, and S. L. Sondhi,
 Phase structure of driven quantum systems,
 \href{http://dx.doi.org/10.1103/PhysRevLett.116.250401}{%
 Phys. Rev. Lett. {\bf 116}, 250401 (2016)}.

\bibitem{else:16}
D. V. Else, B. Bauer, and C. Nayak,
 Floquet time crystals,
 \href{http://dx.doi.org/10.1103/PhysRevLett.117.090402}{%
 Phys. Rev. Lett. {\bf 117}, 090402 (2016)}.

\bibitem{yao:17}
N. Y. Yao, A. C. Potter, I.-D. Potirniche, and A. Vishwanath,
 Discrete time crystals: Rigidity, criticality, and realizations,
 \href{http://dx.doi.org/10.1103/PhysRevLett.118.030401}{%
 Phys. Rev. Lett. {\bf 118}, 030401 (2017)}.

\bibitem{heugel:19}
T. L. Heugel, M. Oscity, A. Eichler, O. Zilberberg, and R. Chitra,
 Classical many-body time crystals,
 \href{http://dx.doi.org/10.1103/PhysRevLett.123.124301}{%
 Phys. Rev. Lett. {\bf 123}, 124301 (2019)}.

\bibitem{kyprianidis:21}
A. Kyprianidis, F. Machado, W. Morong, P. Becker, K. S. Collins, D. V. Else, L.
  Feng, P. W. Hess, C. Nayak, G. Pagano, N. Y. Yao, and C. Monroe,
 Observation of a prethermal discrete time crystal,
 \href{http://dx.doi.org/10.1126/science.abg8102}{%
 Science {\bf 372} (6547), 1192 (2021)}.

\bibitem{guo:13}
L. Guo, M. Marthaler, and G. Sch\"on,
 Phase space crystals: A new way to create a quasienergy band structure,
 \href{http://dx.doi.org/10.1103/PhysRevLett.111.205303}{%
 Phys. Rev. Lett. {\bf 111}, 205303 (2013)}.

\bibitem{guo:16}
L. Guo and M. Marthaler,
 Synthesizing lattice structures in phase space,
 \href{http://dx.doi.org/10.1088/1367-2630/18/2/023006}{%
 New Journal of Physics {\bf 18} (2), 023006 (2016)}.

\bibitem{svensson:18}
I.-M. Svensson, A. Bengtsson, J. Bylander, V. Shumeiko, and P. Delsing,
 Period multiplication in a parametrically driven superconducting resonator,
 \href{http://dx.doi.org/10.1063/1.5026974}{%
 Applied Physics Letters {\bf 113} (2), 022602 (2018)}.

\bibitem{guo:20}
L. Guo and P. Liang,
 Condensed matter physics in time crystals,
 \href{http://dx.doi.org/10.1088/1367-2630/ab9d54}{%
 New Journal of Physics {\bf 22} (7), 075003 (2020)}.

\bibitem{lang:21}
B. Lang and A. D. Armour,
 Multi-photon resonances in josephson junction-cavity circuits,
 \href{http://dx.doi.org/10.1088/1367-2630/abe483}{%
 New Journal of Physics {\bf 23} (3), 033021 (2021)}.

\bibitem{braunstein:87}
S. L. Braunstein and R. I. McLachlan,
 Generalized squeezing,
 \href{http://dx.doi.org/10.1103/PhysRevA.35.1659}{%
 Phys. Rev. A {\bf 35}, 1659 (1987)}.

\bibitem{armour:13}
A. D. Armour, M. P. Blencowe, E. Brahimi, and A. J. Rimberg,
 Universal quantum fluctuations of a cavity mode driven by a {J}osephson
  junction,
 \href{http://dx.doi.org/10.1103/PhysRevLett.111.247001}{%
 Phys. Rev. Lett. {\bf 111}, 247001 (2013)}.

\bibitem{chang:20}
C. W. S. Chang, C. Sab\'{\i}n, P. Forn-D\'{\i}az, F. Quijandr\'{\i}a, A. M.
  Vadiraj, I. Nsanzineza, G. Johansson, and C. M. Wilson,
 Observation of three-photon spontaneous parametric down-conversion in a
  superconducting parametric cavity,
 \href{http://dx.doi.org/10.1103/PhysRevX.10.011011}{%
 Phys. Rev. X {\bf 10}, 011011 (2020)}.

\bibitem{guckenheimer}
J. Guckenheimer and P. Holmes,
 {\em Nonlinear Oscillations, Dynamical Systems, and Bifurcations of Vector
  Fields\/}
 (Springer-Verlag, New York, 1983).

\bibitem{strogatz:00}
S. H. Strogatz,
 {\em Nonlinear Dynamics and Chaos: With Applications to Physics, Biology,
  Chemistry and Engineering\/}
 (Westview Press, 2000).

\bibitem{wustmann:19}
W. Wustmann and V. Shumeiko,
 Parametric effects in circuit quantum electrodynamics,
 \href{http://dx.doi.org/10.1063/1.5116533}{%
 Low Temperature Physics {\bf 45} (8), 848 (2019)}.

\bibitem{arndt:21}
L. Arndt and F. Hassler,
 Universality of photon counting below a local bifurcation threshold,
 \href{http://dx.doi.org/10.1103/PhysRevA.103.023506}{%
 Phys. Rev. A {\bf 103}, 023506 (2021)}.

\bibitem{arnold:89}
V. Arnol'd,
 {\em Mathematical methods of classical mechanics\/}
 (Springer-Verlag, New York, 1989),
 Appendix 7.

\bibitem{tadokoro:20}
Y. Tadokoro, H. Tanaka, and M. I. Dykman,
 Noise-induced switching from a symmetry-protected shallow metastable state,
 \href{http://dx.doi.org/10.1038/s41598-020-66243-y}{%
 Sci Rep {\bf 10}, 10413 (2020)}.

\bibitem{lorch:19}
N. L\"orch, Y. Zhang, C. Bruder, and M. I. Dykman,
 Quantum state preparation for coupled period tripling oscillators,
 \href{http://dx.doi.org/10.1103/PhysRevResearch.1.023023}{%
 Phys. Rev. Research {\bf 1}, 023023 (2019)}.

\bibitem{zhang:17}
Y. Zhang, J. Gosner, S. M. Girvin, J. Ankerhold, and M. I. Dykman,
 Time-translation-symmetry breaking in a driven oscillator: From the quantum
  coherent to the incoherent regime,
 \href{http://dx.doi.org/10.1103/PhysRevA.96.052124}{%
 Phys. Rev. A {\bf 96}, 052124 (2017)}.

\bibitem{zhang:19}
Y. Zhang and M. I. Dykman,
 Nonlocal random walk over floquet states of a dissipative nonlinear
  oscillator,
 \href{http://dx.doi.org/10.1103/PhysRevE.100.052148}{%
 Phys. Rev. E {\bf 100}, 052148 (2019)}.

\bibitem{gosner:20}
J. Gosner, B. Kubala, and J. Ankerhold,
 Relaxation dynamics and dissipative phase transition in quantum oscillators
  with period tripling,
 \href{http://dx.doi.org/10.1103/PhysRevB.101.054501}{%
 Phys. Rev. B {\bf 101}, 054501 (2020)}.

\bibitem{nathan:20}
F. Nathan, G. Refael, M. S. Rudner, and I. Martin,
 Quantum frequency locking and downconversion in a driven qubit-cavity system,
 \href{http://dx.doi.org/10.1103/PhysRevResearch.2.043411}{%
 Phys. Rev. Research {\bf 2}, 043411 (2020)}.

\bibitem{denisenko:16}
M. Denisenko, V. Munyayev, and A. Satanin,
 Quantum fractional resonances in superconducting circuits with an embedded
  josephson junction,
 \href{http://dx.doi.org/10.1088/1742-6596/681/1/012018}{%
 Journal of Physics: Conference Series {\bf 681}, 012018 (2016)}.

\bibitem{svensson:17}
I.-M. Svensson, A. Bengtsson, P. Krantz, J. Bylander, V. Shumeiko, and P.
  Delsing,
 Period-tripling subharmonic oscillations in a driven superconducting
  resonator,
 \href{http://dx.doi.org/10.1103/PhysRevB.96.174503}{%
 Phys. Rev. B {\bf 96}, 174503 (2017)}.

\bibitem{schmid:82}
A. Schmid,
 On a quasiclassical {L}angevin equation,
 \href{http://dx.doi.org/10.1007/BF00681904}{%
 Journal of Low Temperature Physics {\bf 49} (5), 609 (1982)}.
 
\bibitem{Note1}
Here, the tunneling rate can be interpreted as rate constants in the context of rate equations between the different stable fixed-points of the dynamics

\bibitem{kamenev_2011}
A. Kamenev,
 {\em Field Theory of Non-Equilibrium Systems\/}
 (Cambridge University Press, 2011).
 
\bibitem{Note2}
A detailed derivation of the escape rate together with a comparison of the final result to a numerical simulation of the quasi-classical Langevin equation can be found in the supplement.

\bibitem{chinarov:93}
V. A. Chinarov, M. I. Dykman, and V. N. Smelyanskiy,
 Dissipative corrections to escape probabilities of thermal-nonequilibrium
  systems,
 \href{http://dx.doi.org/10.1103/PhysRevE.47.2448}{%
 Phys. Rev. E {\bf 47}, 2448 (1993)}.
 
\bibitem{Note3}
Note that the Poincare cross-section method has been previously employed in Ref.~\cite{tadokoro:20} to obtain the tunneling rate in the classical limit $k_B T\gg\hbar\Omega$.

\end{thebibliography}
\end{document}